\def\be{\begin{equation}}
\def\ee{\end{equation}}
\def\ba{\begin{array}}
\def\ea{\end{array}}
\def\1{{\bf 1}}
\def\p{\prime}
\def\pp{{\prime\prime}}
\def\t{\tau}
\def\R1{{1\!\! 1}}
\def\Rb{{I\!\! R}}
\def\Nb{{I\!\! N}}
\def\Fb{{I\!\! F}}
\def\Cb{\ \hbox{\vrule width 0.6pt height 6pt depth 0pt
		      \hskip -3.5 pt} C}
\documentstyle[12pt]{article}
\begin{document}
\parskip=3pt
\parindent=18pt
\baselineskip=20pt
\setcounter{page}{1}
\centerline{\Large\bf Integrable Lattice Systems}
\vspace{2ex}
\centerline{\Large\bf and Markov Processes}
\vspace{6ex}
\begin{center}
Sergio Albeverio \footnote{ SFB 256; BiBoS; CERFIM(Locarno); Acc. Arch.;
USI(Mendriso)}~~~~
and~~~~
Shao-Ming Fei \footnote{ Institute of Applied Mathematics, Chinese Academy 
of Science, Beijing}\\
Institut f{\"u}r Angewandte Mathematik, Universit{\"a}t Bonn, D-53115, Bonn.\\
\end{center} 
\vskip 1 true cm
\parindent=18pt
\parskip=6pt
\begin{center}
\begin{minipage}{5in}
\vspace{3ex}
\centerline{\large Abstract}
\vspace{4ex}
Lattice systems with certain
Lie algebraic or quantum Lie algebraic symmetries are constructed.
These symmetric models give rise to series of integrable systems. 
As examples the $A_n$-symmetric chain models and the $SU(2)$-invariant
ladder models are investigated. It
is shown that corresponding to these $A_n$-symmetric chain models 
and $SU(2)$-invariant ladder models
there are exactly solvable stationary discrete-time (resp.
continuous-time) Markov chains with
transition matrices (resp. intensity matrices) having spectra which
coincide with the ones of the corresponding integrable models.

\bigskip
\medskip


\end{minipage}
\end{center}

\newpage

\section{Introduction}

Integrable lattice models such as quantum chain and ladder models
have played significant roles
in statistical and condensed matter physics. Many of these models
can be exactly solved in terms of an algebraic or coordinate ``Bethe Ansatz
method" \cite{faddeev}, see e.g., \cite{2345} for chain models with
periodic boundary conditions and fixed boundary
conditions, and \cite{ladder,ladder1} for 2-leg ladder models with open boundary
conditions. The intrinsic symmetry of these integrable chain models 
plays an essential role in finding complete sets of
eigenstates of the systems.

Stochastic models like 
stochastic reaction-diffusion models describing coagulation-decoagulation, 
birth-death processes, pair-creation and 
pair-annihilation of molecules on a chain, have attracted
considerable interest due to their importance in many physical,
chemical and biological processes \cite{evan}. E.g.,
the simplest models for
diffusion-reaction processes describe particles stochastically hopping
on a lattice \cite{spitzer}. These diffusion-reaction models have been studied
in various ways \cite{henkel2}. In particular,
they have been connected to spin-1/2 Heisenberg 
quantum spin chains \cite{alex}, and then
further developed and generalized to $SU(2)$ symmetric spin-$s$ chains \cite{schutz}.
It was shown in \cite{alca} that
the $U_q SU(p/m)$ invariant models \cite{pk} also naturally appear as
time-evolution operators of chemical systems and the $U_q SU(3/0)$,
$U_q SU(1/2)$ and $U_q SU(2/1)$ symmetric chains were discussed
from a similar point of view in \cite{dahmen}. In \cite{henkel1}
an exact solution of a reaction-diffusion process with three-site interactions
(with a special next to nearest neighbour interaction) is presented. 
	 
The stochastic reaction-diffusion systems are studied
in terms of the ``master equation'' which describes the time evolution
of the probability distribution function \cite{alca,4s}. This equation
has the form of a heat equation with potential (a Schr\"odinger equation with
``imaginary time"). For some reaction-diffusion processes the
``Hamiltonians" in the ``master equation" coincide with the generators of
the Hecke algebra \cite{ritten}.
If an integrable system with open boundary condition can be transformed
into a stochastic reaction-diffusion system, e.g., by a unitary
transformation between their respective Hamiltonians, looked upon
as self-adjoint operators acting in the respective Hilbert spaces,
then the stochastic model so obtained is exactly solvable with the same
energy spectrum as the one of the integrable system
\cite{alca,ritten,henkel}.

We have discussed the integrable chain models in \cite{intmod} and
ladder models in \cite{ladder}.
In this paper, we give a systematic description of these models and extend 
the results to general square lattice models having
a certain Lie algebra or quantum Lie algebra symmetry and their
corresponding stationary discrete-time and continuous-time integrable 
stochastic lattice models.

We consider $M \times L$ square lattices and give the construction of 
Lie algebraic (resp. quantum Lie algebraic) invariant lattice models 
in section 2 (in section 3).
In section 4 we discuss the $A_n$ symmetric integrable chain models 
and the $SU(2)$ symmetric integrable ladder models
in the fundamental representation. 
In section 5 we prove that these $A_n$ (resp. $SU(2)$)
symmetric integrable chain (resp. ladder) models can be transformed into both
continuous-time and discrete-time Markov chains.
Some conclusions and remarks are given in section 6.

\smallskip
\section{Lattice Systems with Lie-Algebraic Symmetry}

Let ${\cal A}$ be a bi-algebra with linear operators
multiplication $m$ and coproduct $\Delta$ such that
$m:~{\cal A}\otimes {\cal A}\to {\cal A}$,
$\Delta:~{\cal A}\to {\cal A}\otimes {\cal A}$.
Let ${\bf id}$ denote the identity transformation, 
${\bf id}:~{\cal A}\to {\cal A}$, $p$ the transposition operator,
$p:~{\cal A}\otimes {\cal A}\to {\cal A}\otimes {\cal A}$, 
$p(a\otimes b)=(b\otimes a),~~\forall a,b\in {\cal A}$.
The multiplication $m$ is associative,
$m(m\otimes {\bf id})=m({\bf id}\otimes m)$,
but not commutative in general, $m\circ p\neq m$.
The coproduct operator $\Delta$ is an algebraic homomorphism,
$\Delta(ab)=\Delta(a)\Delta(b),~\forall a,b\in {\cal A}$.
$\Delta(a)$ and $\Delta(b)$ belong to ${\cal A}\otimes {\cal A}$. The
multiplication of tensors is defined by
$(a_1\otimes a_2)(b_1\otimes b_2)=a_1b_1\otimes
a_2b_2$, $a_1,a_2,b_1,b_2\in {\cal A}$.
The coproduct is associative
$(\Delta\otimes{\bf id})\Delta=({\bf id}\otimes\Delta)\Delta$,
but in general not co-commutative $p\circ\Delta\neq\Delta$.
The operation $\Delta$ preserves all the algebraic relations of the algebra
${\cal A}$. It gives a way to find representations of the algebra ${\cal A}$ in the
direct product of spaces.
If a bi-algebra has in addition unit, counit and antipode
operators, it is called a Hopf algebra. Lie algebras are Hopf
algebras with $\Delta$ co-commutative. Quantum algebras are Hopf algebras
that are not co-commutative, see e.g. \cite{pressley} and references 
therein.

A Lie-algebra $A$ is a bi-algebra. Let $e=\{e_\alpha\}$,
$\alpha=1,2,...,n$, be the basis of $A$, satisfying the Lie commutation relations
\be\label{11}
[e_\alpha,e_\beta]=C_{\alpha\beta}^\gamma e_\gamma,
\ee
where $C_{\alpha\beta}^\gamma$ are the structure constants with respect
to the base $e$.

Let $\Delta$ (resp. $C(e)$) be the coproduct operator (resp. Casimir
operator) of the algebra $A$. We have
\be\label{12}
[C(e),e_\alpha]=0,~~~\alpha=1,2,...,n.
\ee
The coproduct operator action on the Lie algebra elements is given by
\be\label{13}
\Delta e_\alpha=e_\alpha\otimes\1+\1\otimes e_\alpha,
\ee
$\1$ stands for the identity operator. It is easy to check that
$$
[\Delta e_\alpha,\Delta e_\beta]=C_{\alpha\beta}^\gamma \Delta e_\gamma.
$$
From the properties of the coproduct, $\Delta C(e)$ is a two-fold tensor
satisfying
\be\label{14}
[\Delta C(e),\Delta e_\alpha]=0,~~~\alpha=1,2,...,n.
\ee

We consider $M\times L$ square lattice systems.
To each point at the $i$-th rung, $i=1,...,L$, and $\theta$-th leg, $\theta =1,...,M$,
of the lattice we associate a (finite dimensional complex) Hilbert space $H_i^\theta$.
Let $H_i=H_i^1 \otimes H_i^2\otimes...\otimes H_{i}^M$.
We can then associate to the whole lattice
the tensor product $H_1\otimes H_2\otimes...\otimes H_{L}$.
The generators of the algebra $A$ acting on this Hilbert space
associated with the above lattice are given by
$E_\alpha=\Delta^{ML-1} e_\alpha$, $\alpha=1,2,...,n$,
where we have defined
\be\label{deltam}
\Delta^m=(\underbrace{{\1}\otimes ... \otimes{\1}}_{m~times}\otimes\Delta)
...({\1}\otimes{\1}\otimes\Delta)
({\1}\otimes\Delta)\Delta,~~\forall\, m\in\Nb.
\ee
$E_\alpha$ also generates the Lie algebra $A$:
$[E_\alpha,E_\beta]=C_{\alpha\beta}^\gamma E_\gamma$.

Let $\Delta_n^m$ be an $m$-fold tensor operator with operator $\Delta$ on the $n$-th,
$1\leq n\leq m$, tensor space and identity on the rest. For instance,
$\Delta_1^1=\Delta$, $\Delta_1^2=\Delta\otimes\1$, $\Delta_2^2=\1\otimes\Delta$,
$\Delta_1^3=\Delta\otimes\1\otimes\1$, $\Delta_2^3=\1\otimes\Delta\otimes\1$,
$\Delta_3^3=\1\otimes\1\otimes\Delta$. Set
\be\label{h}
h=\sum_{i_M=1}^M...\sum_{i_2=1}^2 \sum_{i_1=1}^1 a_{i_1 i_2...i_M}
\Delta_{i_M}^M...\Delta_{i_2}^2\Delta_{i_1}^1 C(e),
\ee
where $a_{i_1 i_2...i_M}\in\Cb$ such that $h$ is hermitian.
Let $\Fb$ denote a real entire function defined on the $ML$-fold tensor space 
$A\otimes A\otimes...\otimes A$ of the algebra $A$. We call 
\be\label{18}
H=\sum_{i=1}^{L-1} \Fb(h)_{i,i+1}
\ee
the (quantum mechanics) Hamiltonian associated with the lattice.
Here $\Fb(h)_{i,i+1}$ means that the $2M$-fold tensor
element $\Fb(h)$ is associated with the $i$ and $i+1$-th rungs (columns) of the
lattice and acts on the space $H_i \otimes H_{i+1}$, i.e.,
\be\label{pp}
\Fb (h)_{i,i+1}=\1_1 \otimes...\otimes\1_{i-1}\otimes \Fb(h)
\otimes\1_{i+2}\otimes... \otimes\1_{L},
\ee
where $\1_i=\1_i^1 \otimes ... \otimes\1_i^M$ is the identity operator acting on the $i$-th
rungs of the lattice.

{\sf [Theorem 1]}. The Hamiltonian $H$ is a self-adjoint operator acting
in $H_1\otimes H_2\otimes...\otimes H_{L}$ and is invariant under the algebra $A$.
	
{[\sf Proof].} That $H$ is self-adjoint is immediate from the definition.
To prove the invariance of $H$
it suffices to prove $[H,E_\alpha]=0$, $\alpha=1,2,...,n$.

From the formula for the coproduct we have
\be\label{19}
E_\alpha=\sum_{i=1}^{L-1}(e_{\alpha})_i,
\ee
where $(e_{\alpha})_i=\1_1\otimes...\otimes\1_{i-1}\otimes
\Delta^{M-2}e_\alpha\otimes\1_{i+1}\otimes... \otimes\1_{L}$,
$\Delta^{M-2}$ as defined in (\ref{deltam}). For $M=1$, $\Delta^{-1}$ is understood as
identity operator. 

From (\ref{14}) it is direct to prove that
$[h,\Delta^{2M-2} e_{\alpha}]=0$.
Obviously $[\Fb(h)_{i,i+1},(e_{\alpha})_j]=0$, $\forall j\neq i,i+1$.
Therefore we have, for all
$\alpha=1,2,...,n$:
\be\label{he}
\ba{rcl}
[H,E_\alpha]&=&\displaystyle\left[\sum_{i=1}^{L-1}\Fb(h)_{i,i+1},
\displaystyle\sum_{j=1}^{i-1}(e_{\alpha})_j+
\displaystyle\sum_{k=i+2}^{L-1}(e_{\alpha})_k
+(e_{\alpha})_i+(e_{\alpha})_{i+1}\right]\\[5mm]
&=&\displaystyle\sum_{i=1}^{L-1}\left[\Fb(h)_{i,i+1},
(e_{\alpha})_i+(e_{\alpha})_{i+1}\right]
=\displaystyle\sum_{i=1}^{L-1}\left[
\Fb(h)_{i,i+1},(\Delta^{2M-2} e_\alpha)_{i,i+1}\right]=0.
\ea
\ee
\hfill $\rule{3mm}{3mm}$

\smallskip
\section{\bf Lattice Models with Quantum Lie Algebraic Symmetry}

Let $e=\{e_\alpha,f_\alpha,h_\alpha\}$, $\alpha=1,2,...,n$, 
be the Chevalley basis of a Lie algebra $A$ with rank $n$.
Let $e^\p=\{e_\alpha^\p,f_\alpha^\p,h_\alpha^\p\}$, 
$\alpha=1,2,...,n$, be the corresponding elements of the
quantum (q-deformed) Lie algebra  $A_q$. We denote by $r_\alpha$ the simple
roots of the Lie algebra $A$.
The quantum algebra generated by $\{e_\alpha^\p,f_\alpha^\p,h_\alpha^\p\}$
is defined by the following relations \cite{pressley}:
\be\label{23}
\ba{rcl}
[h_\alpha^\p,h_\beta^\p]&=&0,~~~~
[h_\alpha^\p,e_\beta^\p]=a_{\alpha\beta}e_\beta^\p,\\[4mm]
[h_\alpha^\p,f_\beta^\p]&=&-a_{\alpha\beta}f_\beta^\p,~~~~
[e_\alpha^\p,f_\beta^\p]=\delta_{\alpha,\beta}
\displaystyle\frac{q^{d_\alpha h_\alpha^\p}-q^{-d_\alpha h_\alpha^\p}}
{q^{d_\alpha}-q^{-d_\alpha}}
\ea
\ee
together with the quantum Serre relations
\be\label{24}
\ba{l}
\displaystyle\sum_{\gamma=0}^{1-a_{\alpha\beta}}(-1)^\gamma
\left[\ba{c}1-a_{\alpha\beta}\\[1mm]\gamma\ea\right]_{q^{d_\alpha}}
(e_\alpha^\p)^\gamma e_\beta^\p(e_\alpha^\p)^{1-a_{\alpha\beta}-
\gamma}=0,~~~i\neq j, \\[6mm]
\displaystyle\sum_{\gamma=0}^{1-a_{\alpha\beta}}(-1)^\gamma
\left[\ba{c}1-a_{\alpha\beta}\\[1mm]\gamma\ea\right]_{q^{d_\alpha}}
(f_\alpha^\p)^\gamma f_\beta^\p(f_\alpha^\p)^{1-a_{\alpha\beta}-
\gamma}=0,~~~i\neq j,
\ea
\ee
where for $m\geq n\in\Nb$,
$$
\left[\ba{c}m\\[1mm]n\ea\right]_{q}=\displaystyle\frac{[m]_q!}
{[n]_q![m-n]_q!},~~~~
[n]_q!=[n]_q[n-1]_q...[2]_q[1]_q,~~~~
[n]_q=\displaystyle\frac{q^n-q^{-n}}{q-q^{-1}},
$$
$(a_{\alpha\beta})$ is the Cartan matrix,
$$
a_{\alpha\beta}=\frac{1}{d_\alpha}(r_\alpha\cdot r_\beta),~~~
d_\alpha=\frac{1}{2}(r_\alpha\cdot r_\alpha),
$$
$q$ is a complex quantum parameter such that $q^{d_\alpha}\neq\pm1,0$. 

The coproduct operator $\Delta^\prime$ of the quantum algebra $A_q$ is
given by
\begin{eqnarray}
\Delta^\p h_\alpha^\p&=&h_\alpha^\p\otimes\1+\1\otimes h_\alpha^\p,\label{25}\\[3mm]
\Delta^\p e_\alpha^\p&=&e_\alpha^\p\otimes q^{-d_\alpha h_\alpha^\p}
+q^{d_\alpha h_\alpha^\p}\otimes e_\alpha^\p,\label{26}\\[3mm]
\Delta^\p f_\alpha^\p&=&f_\alpha^\p\otimes q^{-d_\alpha h_\alpha^\p}
+q^{d_\alpha h_\alpha^\p}\otimes f_\alpha^\p.\label{27}
\end{eqnarray}
It is straightforward to check that $\Delta^\p$ preserves all the algebraic
relations in (\ref{23}) and (\ref{24}).

Let $C_q(e^\prime)$ be the Casimir operator of $A_q$, i.e.,
$[C_q(e^\prime), a]=0,~\forall a\in A_q$. For any entire function $\Fb$ 
of $C_q(e^\prime)$, we have
\be\label{28}
[\Fb(C_q(e^\prime)), a]=0,~\forall a\in A_q
\ee
and
\be\label{29}
[\Delta^\p \Fb(C_q(e^\prime)), \Delta^\p a]=0,~\forall a\in A_q.
\ee
Especially, by formula (\ref{25}) one gets
\be\label{30}
\Delta^\p q^{\pm d_\alpha h_\alpha^\p}=q^{\pm d_\alpha h_\alpha^\p}\otimes 
q^{\pm d_\alpha h_\alpha^\p}.
\ee
Hence
\be\label{31}
[\Delta^\p \Fb(C_q(e^\prime)), \Delta^\p q^{\pm d_\alpha h_\alpha^\p}]
=[\Delta^\p \Fb(C_q(e^\prime)), q^{\pm d_\alpha h_\alpha^\p}\otimes 
q^{\pm d_\alpha h_\alpha^\p}]=0.
\ee

{\sf [Theorem 2]}. The lattice model defined by the following Hamiltonian
acting in $H_1 \otimes H_2\otimes...\otimes H_{L}$ is 
invariant under the quantum algebra $A_q$:
\be\label{33}
H_q=\sum_{i=1}^{L-1}\Fb(h_q)_{i,i+1},
\ee
where 
$$h_q=\sum_{i_M=1}^M...\sum_{i_2=1}^2 \sum_{i_1=1}^1 a_{i_1 i_2...i_M}
\Delta_{i_M}^{\p\, M}...\Delta_{i_2}^{\p\, 2} \Delta_{i_1}^{\p\, 1} C_q(e^\p),
$$
with $a_{i_1 i_2...i_M}\in\Cb$ such that $h_q$ is hermitian.

{\sf [Proof]}. The generators of $A_q$ on the lattice are given by
\be\label{32}
\ba{rcl}
H^\p_\alpha&=&\Delta^{\p\, ML-2} h^\p_\alpha=
\displaystyle\sum_{i=1}^{L-1}
\1_1\otimes...\otimes\1_{i-1}\otimes
(\Delta^{\p\, 2M-2}h_\alpha^\p)_{i,i+1}
\otimes\1_{i+2}\otimes... \otimes\1_{L},\\[4mm]
E^\p_\alpha&=&\displaystyle\sum_{i=1}^{L-1}
q^{d_\alpha h^\p_\alpha}\otimes...\otimes q^{d_\alpha h^\p_\alpha}
\otimes (\Delta^{\p\, 2M-2}e_\alpha^\p)_{i,i+1}\otimes q^{-d_\alpha h^\p_\alpha}\otimes...
\otimes q^{-d_\alpha h^\p_\alpha},\\[6mm]
F^\p_\alpha&=&\displaystyle\sum_{i=1}^{L-1}
q^{d_\alpha h^\p_\alpha}\otimes...\otimes q^{d_\alpha h^\p_\alpha}
\otimes (\Delta^{\p\, 2M-2}f_\alpha^\p)_{i,i+1}\otimes q^{-d_\alpha h^\p_\alpha}\otimes...
\otimes q^{-d_\alpha h^\p_\alpha}.
\ea
\ee

From (\ref{29}) and (\ref{30}) we have
$[h_q,\Delta^{\p\,2M-2}h_\alpha^\p]=[h_q,\Delta^{\p\,2M-2}e_\alpha^\p]
=[h_q,\Delta^{\p\,2M-2}f_\alpha^\p]=0$. Therefore
$$
\ba{rcl}
[H_q,E^\p_\alpha]
&=&\displaystyle\left[\sum_{i=1}^{L-1}
\Fb(h_q)_{i,i+1},\left(\sum_{j=1}^{i-2}+\sum_{j=i+2}^{L-1}\right)
\left(q^{d_\alpha h^\p_\alpha}\otimes...
\otimes (\Delta^{\p\, 2M-2}e_\alpha^\p)_{j,j+1}\otimes...
\otimes q^{-d_\alpha h^\p_\alpha}\right)\right.\\[6mm]
&&\left.+ q^{d_\alpha h^\p_\alpha}\otimes...
\otimes (\Delta^{\p\, 2M-2}e_\alpha^\p)_{i,i+1}\otimes...
\otimes q^{-d_\alpha h^\p_\alpha}
\right]\\[4mm]
&=&\displaystyle\sum_{i=1}^{L-1}\left[\Fb(h_q),\Delta^{\p\,2M-2} (e^\p_\alpha)\right]_{i,i+1}=0.
\ea
$$
$[H_q,F^\p_\alpha]=0$ is obtained similarly.
$[H_q,H^\p_\alpha]=0$ can be proved like (\ref{he}).
Hence $H_q$ commutes with the generators of $A_q$.
\hfill $\rule{3mm}{3mm}$

The Hamiltonian system (\ref{33}) is expressed by the quantum
algebraic generators $e^\p=(h^\p_\alpha,e^\p_\alpha,f^\p_\alpha)$. 
Assume now that $e\to e^\prime(e)$ is an
algebraic map from $A$ to $A_q$ (we remark that for algebras with
three generators like $A_1$, both classical and quantum
algebraic maps can be discussed in terms of the two dimensional
manifolds related to the algebras, see  \cite{fa}). We then have
\be\label{35}
H_q=\sum_{i=1}^{L-1} \Fb(h_q(e^\p(e))_{i,i+1}.
\ee
In this way we obtain lattice models having quantum algebraic symmetry
but expressed in terms of the usual Lie algebraic generators $\{e_{\alpha}\}$
with manifest physical meanings.

\smallskip
\section{Integrable Lattice Models with Lie algebraic Symmetry}

\subsection{\bf Quantum Yang-Baxter Equation}

The quantum Yang-Baxter equation (QYBE) \cite{yang} is the ``master 
equation" for integrable models in statistical mechanics.
It plays an important role in a variety of problems
in theoretical physics such as the study of 
exactly solvable models like the six and
eight vertex models in statistical mechanics \cite{baxter}, 
of integrable model field theories \cite{3}, of exact S-matrix theoretical
models \cite{4}, as well as in the investigation of
two dimensional field theories involving fields with 
intermediate statistics \cite{5}, in conformal field theory
and in the study of quantum groups \cite{pressley}. 
In the following we investigate the integrability of lattice 
models having a Lie algebraic 
symmetry constructed in section 2. We also
present a series of solutions of the QYBE from the construction of
integrable models.

Let $V$ be a complex vector space and $R$ the solution of QYBE without
spectral parameters, see e.g. \cite{pressley}. Then $R$ 
takes values in $End_{\Cb}(V\otimes V)$. The QYBE is
\begin{equation}\label{36}
R_{12}R_{13}R_{23}=R_{23}R_{13}R_{12}.
\end{equation}
Here $R_{ij}$ denotes the matrix on the complex vector space 
$V\otimes V\otimes V$, acting as $R$ on the $i$-th and the 
$j$-th components and as the identity on the other components.

Let $\check{R}=Rp$ ($p$ is the transposition operator).
Then the QYBE (\ref{36}) becomes
\be\label{37}
\check{R}_{12}\check{R}_{23}\check{R}_{12}=
\check{R}_{23}\check{R}_{12}\check{R}_{23},
\ee
where  $\check{R}_{12}=\check{R}\otimes\1_V$, $\check{R}_{23}=\1_V\otimes
\check{R}$ and $\1_V$ is the identity operator on $V$. 

In the following we say that a lattice model with nearest neighbours
interactions having a (quantum mechanical) Hamiltonian of the form
\be\label{38}
H=\sum_{i=1}^{L-1}({\cal H})_{i,i+1}
\ee
is integrable in the sense that the operator
${\cal H}$ satisfies the QYBE relation (\ref{37}), i.e.,
\be\label{39}
({\cal H})_{12}({\cal H})_{23}({\cal H})_{12}=
({\cal H})_{23}({\cal H})_{12}({\cal H})_{23},
\ee
where $({\cal H})_{12}={\cal H}\otimes \1_V$ and $({\cal H})_{23}=\1_V
\otimes {\cal H}$. Here ${\cal H}$ is a solution of the Yang-Baxter
equation without spectral parameters. After ``baxterization" the
Hamiltonian system (\ref{38}) satisfying relation (\ref{39}) can in
principle be exactly solved in terms of algebraic Bethe Ansatz method, see e.g.
\cite{faddeev}. Here the vector space $V$ is taken to be the Hilbert spaces
associated with one rung of the lattice.

\smallskip
\subsection{\bf Integrable $A_n$ Symmetric Chain Models}

The integrability of the models having a Lie algebraic symmetry
presented in section 2 depends on the detailed representation of the
corresponding symmetry algebra. In this section we investigate the integrability 
of chain models with nearest neighbours interactions and Lie
algebraic symmetry $A_n$.

Let $(a_{\alpha\beta})$ be the Cartan matrix of the $A_n$ algebra. In
the Chevalley basis the algebra $A_n$ is spanned by the generators 
$\{h_\alpha,e_\alpha,f_\alpha\}$, $\alpha=1,2,...,n$, with the following
algebraic relations:
\be\label{40}
[h_\alpha,h_\beta]=0,~~~
[h_\alpha,e_\beta]=a_{\alpha\beta}e_\beta,~~~
[h_\alpha,f_\beta]=-a_{\alpha\beta}f_\beta,~~~
[e_\alpha,f_\beta]=\delta_{\alpha\beta}h_\alpha,
\ee
together with the generators with respect to non simple roots,
\be\label{41}
e_{\alpha...\beta\gamma}=[e_\alpha,...,[e_\beta,e_\gamma]...],~~~
f_{\alpha...\beta\gamma}=[f_\alpha,...,[f_\beta,f_\gamma]...].
\ee

Let $E_{\alpha\beta}$ be an $(n+1)\times (n+1)$ matrix such that 
$(E_{\alpha\beta})_{\gamma\delta}=\delta_{\alpha\gamma}\delta_{\beta\delta}$,
i.e., the only non zero element of the matrix $E_{\alpha\beta}$ is $1$
at row $\alpha$ and column $\beta$. Hence
\be\label{42}
E_{\alpha\beta}E_{\gamma\delta}
=\delta_{\beta\gamma}E_{\alpha\delta},~~~~~
[E_{\alpha\beta},E_{\gamma\delta}]=\delta_{\beta\gamma}E_{\alpha\delta}
-\delta_{\delta\alpha}E_{\beta\gamma}.
\ee

For the fundamental representation we take the basis of the algebra $A_n$ as
\be\label{43}
\ba{l}
h_\alpha =E_{\alpha\alpha}-E_{\alpha+1,\alpha+1},~~~\alpha=1,2,...,n\\[3mm]
\left.\ba{l}
e=\{E_{\alpha\beta}\}\\[3mm]
f=\{E_{\beta\alpha}\}
\ea
~~\right\}~~~~\beta>\alpha=1,2,...,n
\ea
\ee
Both $\{e_\alpha\}$ and $\{f_\alpha\}$ have a total of $n(n+1)/2$ generators.

With respect to the basis (\ref{43}), the Casimir operator of the 
algebra $A_n$ is given by
\be\label{44}
\ba{rcl}
C_{A_n}&=&(n+1)\displaystyle\sum_{\alpha=1}^{n(n+1)/2}
(e_\alpha f_\alpha+f_\alpha e_\alpha)
+\displaystyle\sum_{\alpha=1}^{n}\alpha(n+1-\alpha)h_\alpha^2\\[5mm]
&&+\displaystyle\sum_{\alpha=1}^{n}\displaystyle\sum_{\beta=1}^{n-\alpha}
2\alpha(n+1-\alpha-\beta)h_\alpha h_{\alpha+\beta}-a,
\ea
\ee
where $a$ is an arbitrary real constant.

The coproduct operator $\Delta$ is given by
\be\label{45}
\ba{ll}
\,~\Delta(\1)=\1\otimes\1,~~~~
\Delta(h_\alpha)=h_\alpha\otimes\1+\1\otimes h_\alpha,
&~~~\alpha=1,2,...,n\\[3mm]
\left.\ba{l}
\Delta(e_\beta)=e_\beta\otimes\1+\1\otimes e_\beta\\[3mm]
\Delta(f_\beta)=f_\beta\otimes\1+\1\otimes f_\beta
\ea
~~\right\}&~~~~\beta=1,2,...,n(n+1)/2,
\ea
\ee
where the identity operator $\1$ is the $(n+1)\times (n+1)$ identity
matrix.

By (\ref{44}) and (\ref{45}) we have
\be\label{46}
\ba{ll}
\Delta C_{A_n}=&C_{A_n}\otimes\1 +\1\otimes C_{A_n}-a\1\otimes\1 \\[4mm]
&+(n+1)\displaystyle\sum_{\alpha=1}^{n(n+1)/2}
(e_\alpha\otimes f_\alpha+f_\alpha\otimes e_\alpha)
+\displaystyle\sum_{\alpha=1}^{n}\alpha(n+1-\alpha)h_\alpha\otimes h_\alpha\\[5mm]
&+\displaystyle\sum_{\alpha=1}^{n}\displaystyle\sum_{\beta=1}^{n-\alpha}
\alpha(n+1-\alpha-\beta)(h_\alpha\otimes h_{\alpha+\beta}
+h_{\alpha+\beta}\otimes h_{\alpha}).
\ea
\ee

It is easy to check that under the representation (\ref{43}) $C_{A_n}$
is equal to $n(n+2)\1$. Therefore the
sum of the first two terms on the right hand side of (\ref{46}) is
$2n(n+2)\1\times\1$. In the following we take $a$ in (\ref{46}) to be
$2n(n+2)$ so that the terms that are
proportional to the $(n+1)^2\times (n+1)^2$ identity matrix 
will disappear in (\ref{46}).

From (\ref{43}) and (\ref{46}) we have
\be\label{47}
\ba{ll}
\Delta C_{A_n}=&
(n+1)\displaystyle\sum_{\alpha\neq\beta=1}^{n+1}E_{\alpha\beta}\otimes
E_{\beta\alpha}\\[5mm]
&+\displaystyle\sum_{\alpha=1}^{n}\alpha(n+1-\alpha)
(E_{\alpha\alpha}-E_{\alpha+1,\alpha+1})\otimes 
(E_{\alpha\alpha}-E_{\alpha+1,\alpha+1})\\[5mm]
&+\displaystyle\sum_{\alpha=1}^{n}\displaystyle\sum_{\beta=1}^{n-\alpha}
\alpha(n+1-\alpha-\beta)[(E_{\alpha\alpha}-E_{\alpha+1,\alpha+1})
\otimes (E_{\alpha+\beta,\alpha+\beta}-
E_{\alpha+\beta+1,\alpha+\beta+1})\\[5mm]
&+ (E_{\alpha+\beta,\alpha+\beta}-E_{\alpha+\beta+1,\alpha+\beta+1})
\otimes (E_{\alpha\alpha}-E_{\alpha+1,\alpha+1})].
\ea
\ee

$\Delta C_{A_n}$ in (\ref{47}) is an $(n+1)^2\times (n+1)^2$ matrix. Its
matrix representation is
\be\label{48}
\ba{ll}
(\Delta C_{A_n})_{\alpha\beta}=
&\delta_{\alpha\beta}[(n+1)\delta_{\alpha,l(n+1)+l+1}-1]\\[3mm]
&+(n+1)[\delta_{\alpha,j(n+2)+k+2}\delta_{\beta,(j+1)(n+2)+k(n+1)}\\[3mm]
&+\delta_{\beta,j(n+2)+k+2}\delta_{\alpha,(j+1)(n+2)+k(n+1)}],
\ea
\ee
where $\alpha,\beta=1,2,...,(n+1)^2$, $l=0,1,...,n$, $j=0,1,...,n-1$, 
$k=0,1,...,n-j-1$, $\delta_{\alpha,j(n+2)+k+2}=0$ if 
$\alpha\neq j(n+2)+k+2$ for all possible values of $j$ and $k$. 
For example,
\be\label{49}
\Delta C_{A_1}=\left(
\ba{cccc}
1&0&0&0\\[2mm]
0&-1&2&0\\[2mm]
0&2&-1&0\\[2mm]
0&0&0&1
\ea
\right).
\ee

{\sf [Lemma 1].} $\Delta C_{A_n}$ satisfies the following relation
\be\label{53}
(\Delta C_{A_n})^2+2\Delta C_{A_n}-n(n+2)\1\otimes\1=0.
\ee

{\sf [Proof].} From (\ref{48}) we have
$$
\ba{rcl}
[(\Delta C_{A_n})^2]_{\alpha\gamma}
&=&\displaystyle\sum_{\beta=1}^{(n+1)^2}
(\Delta C_{A_n})_{\alpha\beta}(\Delta C_{A_n})_{\beta\gamma}\\[3mm]
&=&\delta_{\alpha\gamma}[(n+1)^2\delta_{\alpha,l(n+1)+l+1}
\delta_{\gamma,l^\p (n+1)+l^\p +1}]\\[3mm]
&&-(n+1)(\delta_{\alpha,l(n+1)+l+1}+\delta_{\gamma,l^\p (n+1)+l^\p +1})+1]
\\[3mm]
&&-2(n+1)[\delta_{\alpha,j(n+2)+k+2}\delta_{\gamma,(j+1)(n+2)+k(n+1)}
+\delta_{\gamma,j(n+2)+k+2}\delta_{\alpha,(j+1)(n+2)+k(n+1)}]\\[3mm]
&&+(n+1)^2[\delta_{\alpha,j(n+2)+k+2}\delta_{\gamma,j(n+2)+k+2}
+\delta_{\alpha,(j+1)(n+2)+k(n+1)}\delta_{\gamma,(j+1)(n+2)+k(n+1)}]\\[3mm]
&=&-2(\Delta C_{A_n})_{\alpha\gamma}+
(n+1)^2\delta_{\alpha\gamma}
[\delta_{\alpha,j(n+2)+k+2}+\delta_{\alpha,(j+1)(n+2)+k(n+1)}]\\[3mm]
&&+(n+1)^2\delta_{\alpha\gamma}\delta_{\alpha,l(n+1)+l+1}-\delta_{\alpha\gamma}
\\[3mm]
&=&-2(\Delta C_{A_n})_{\alpha\gamma}+n(n+2)\delta_{\alpha\gamma},
\ea
$$
where the identity
\be\label{identity}
\delta_{\alpha,l(n+1)+l+1}+
\delta_{\alpha,j(n+2)+k+2}+\delta_{\alpha,(j+1)(n+2)+k(n+1)}=1,
\ee
$l=0,1,...,n$, $j=0,1,...,n-1$, $k=0,1,...,n-j-1$, has been used.
\hfill $\rule{3mm}{3mm}$

{\sf [Lemma 2]}. The coproduct of the $A_n$ Casimir operator $\Delta
C_{A_n}$ has the following properties:
\be\label{l21}
\ba{l}
(\Delta C_{A_n}\otimes\1)(\1\otimes\Delta C_{A_n})(\Delta C_{A_n}\otimes\1)
\\[3mm]
-n[(\1\otimes\Delta C_{A_n})(\Delta C_{A_n}\otimes\1)+
(\Delta C_{A_n}\otimes\1)(\1\otimes\Delta C_{A_n})]\\[3mm]
+(n^2-1)(\Delta C_{A_n}\otimes\1)+n^2(\1\otimes\Delta C_{A_n})
+n(1-n^2)\1\otimes\1\otimes\1=0
\ea
\ee
and
\be\label{l22}
\ba{l}
(\1\otimes\Delta C_{A_n})(\Delta C_{A_n}\otimes\1)(\1\otimes\Delta C_{A_n})
\\[3mm]
-n[(\1\otimes\Delta C_{A_n})(\Delta C_{A_n}\otimes\1)+
(\Delta C_{A_n}\otimes\1)(\1\otimes\Delta C_{A_n})]\\[3mm]
+(n^2-1)(\1\otimes\Delta C_{A_n})+n^2(\Delta C_{A_n}\otimes\1)
+n(1-n^2)\1\otimes\1\otimes\1=0.
\ea
\ee

{\sf [Proof]}. By using the representation of $\Delta C_{A_n}$ in (\ref{48})
we have
\be\label{dc1}
\ba{rcl}
(\Delta C_{A_n}\otimes\1)_{\alpha\beta}
&=&(\Delta C_{A_n})_{(\alpha-\gamma)/(n+1)+1,(\beta-\gamma)/(n+1)+1}\\[3mm]
&=&\delta_{\alpha\beta}[(n+1)\delta_{\alpha-\gamma,l(n+1)(n+2)}-1]\\[3mm]
&&+(n+1)[\delta_{\alpha-\gamma,(n+1)(j(n+2)+k+1))}\delta_{\beta-\gamma,
(n+1)(j(n+2)+(k+1)(n+1))}\\[3mm]
&&+\delta_{\beta-\gamma,(n+1)(j(n+2)+k+1)}\delta_{\alpha-
\gamma,(n+1)(j(n+2)+(k+1)(n+1))}]
\ea
\ee
and
\be\label{dc2}
\ba{rcl}
(\1\otimes\Delta C_{A_n})_{\alpha\beta}
&=&(\Delta C_{A_n})_{\alpha-(n+1)^2(\gamma^\p-1),\beta-(n+1)^2(\gamma^\p-1)}\\[3mm]
&=&\delta_{\alpha\beta}[(n+1)\delta_{\alpha-(n+1)^2(\gamma^\p-1),l(n+1)+l+1)}-1]\\[3mm]
&&+(n+1)[\delta_{\alpha-(n+1)^2(\gamma^\p-1),j(n+2)+k+2))}\delta_{\beta-(n+1)^2(\gamma^\p-1),
(j+1)(n+2)+k(n+1)}\\[3mm]
&&+\delta_{\beta-(n+1)^2(\gamma^\p-1),j(n+2)+k+2}
\delta_{\alpha-(n+1)^2(\gamma^\p-1),(j+1)(n+2)+k(n+1)}],
\ea
\ee
where $\alpha,\beta=1,...,(n+1)^3$, $l=0,1,...,n$, $j=0,1,...,n-1$ and
$k=0,1,...,n-j-1$ as in formula (\ref{48}), $\gamma=1,...,n+1$ such that 
$(\alpha-\gamma)/(n+1)$ and $(\beta-\gamma)/(n+1)$ in (\ref{dc1}) are integers
and $\gamma^\p=1,...,n+1$ in (\ref{dc2}).

Using the formulae (\ref{dc1}) and (\ref{dc2}) one can get (\ref{l21}) and
(\ref{l22}) from straightforward calculations. \hfill $\rule{3mm}{3mm}$

From Theorem 1 we know that the following Hamiltonian is invariant 
under $A_{n}$
\be\label{54}
H=\sum_{i=1}^{L-1}\Fb(\Delta C_{A_n})_{i,i+1}.
\ee
For the given representation (\ref{43}) of $A_n$ the integrability of
(\ref{48}) depends on the form of the entire function $\Fb$. Due to the
relation (\ref{53}) in {\sf Lemma 1}, $(\Delta C_{A_n})^l$, 
$l\geq 2$, can be expressed as $c \Delta C_{A_n}+c^\p\1\otimes\1$ for
some real constants $c$ and $c^\p$. Therefore $\Fb(\Delta C_{A_n})$ 
is a polynomial in $\Delta C_{A_n}$ up to powers of order one.

{\sf [Theorem 3]}. The following $A_n$ invariant Hamiltonian
is integrable

\be\label{55}
\ba{rcl}
H_{A_n}&=&\displaystyle\sum_{i=1}^{L-1}({\cal H})_{i,i+1}
=\displaystyle\sum_{i=1}^{L-1}(\Delta C_{A_n}+1)_{i,i+1}\\[4mm]
&=&\displaystyle
\sum_{i=1}^{L-1}\left[(n+1)\displaystyle\sum_{\alpha=1}^{n(n+1)/2}
((e_\alpha)_i(f_\alpha)_{i+1}+(f_\alpha)_i(e_\alpha)_{i+1})
+\displaystyle\sum_{\alpha=1}^{n}\alpha(n+1-\alpha)
(h_\alpha)_i(h_\alpha)_{i+1}\right.\\[5mm]
&&\left.+\displaystyle\sum_{\alpha=1}^{n}\displaystyle\sum_{\beta=1}^{n-\alpha}
\alpha(n+1-\alpha-\beta)((h_\alpha)_i(h_{\alpha+\beta})_{i+1}
+(h_{\alpha+\beta})_i(h_{\alpha})_{i+1})\right]+L-1,
\ea
\ee
where ${\cal H}=\Delta C_{A_n}+1$ and the number $1$ should be 
understood as the identity operator, $\1\otimes \1$, on the tensor space
$H_1\otimes...\otimes H_{L}$. 

{\sf [Proof]}. What we have to prove is that ${\cal H}$ satisfies the
QYBE (\ref{39}), i.e., $({\cal H})_{12}({\cal H})_{23}({\cal H})_{12}=
({\cal H})_{23}({\cal H})_{12}({\cal H})_{23}$,
where $({\cal H})_{12}=(\Delta C_{A_n}+1)\otimes \1$ and
$({\cal H})_{23}=\1 \otimes(\Delta C_{A_n}+1)$. We have
$$
\ba{rcl}
({\cal H})_{12}({\cal H})_{23}({\cal H})_{12}
&=&(\1\otimes\Delta C_{A_n})(\Delta C_{A_n}\otimes\1)+
(\Delta C_{A_n}\otimes\1)(\Delta C_{A_n}\otimes\1)\\[4mm]
&&+(\Delta C_{A_n}\otimes\1)(\1\otimes\Delta C_{A_n})
(\Delta C_{A_n}\otimes\1)
+(\Delta C_{A_n}\otimes\1)(\1\otimes\Delta C_{A_n})\\[4mm]
&&+2\Delta C_{A_n}\otimes\1+\1\otimes\Delta C_{A_n}
+\1\otimes\1\otimes\1
\ea
$$
and
$$
\ba{rcl}
({\cal H})_{23}({\cal H})_{12}({\cal H})_{23}
&=&(\1\otimes\Delta C_{A_n})(\1\otimes\Delta C_{A_n})
+(\1\otimes\Delta C_{A_n})(\Delta C_{A_n}\otimes\1)\\[4mm]
&&+(\1\otimes\Delta C_{A_n})(\Delta C_{A_n}\otimes\1)
(\1\otimes\Delta C_{A_n})
+(\Delta C_{A_n}\otimes\1)(\1\otimes\Delta C_{A_n})\\[4mm]
&&+2(\1\otimes\Delta C_{A_n})+(\Delta C_{A_n}\otimes\1)
+\1\otimes\1\otimes\1.
\ea
$$
Hence
\be\label{57}
({\cal H})_{12}({\cal H})_{23}({\cal H})_{12}-
({\cal H})_{23}({\cal H})_{12}({\cal H})_{23}=
I+II+III,
\ee
where
$$
\ba{rcl}
I&=&(\Delta C_{A_n}\otimes\1)(\Delta C_{A_n}\otimes\1)
-(\1\otimes\Delta C_{A_n})(\1\otimes\Delta C_{A_n}),\\[4mm]
II&=&(\Delta C_{A_n}\otimes\1)(\1\otimes\Delta C_{A_n})(\Delta C_{A_n}\otimes\1)
-(\1\otimes\Delta C_{A_n})(\Delta C_{A_n}\otimes\1)(\1\otimes\Delta
C_{A_n}),\\[4mm]
III&=&\Delta C_{A_n}\otimes\1-\1\otimes\Delta C_{A_n}.
\ea
$$

Using (\ref{53}) we have
$$
\ba{rcl}
I&=&(\Delta C_{A_n})^2\otimes\1-\1\otimes (\Delta C_{A_n})^2\\[4mm]
&=&-(2\Delta C_{A_n}-n(n+2)\1\otimes\1)\otimes\1
+\1\otimes (2\Delta C_{A_n}-n(n+2)\1\otimes\1)\\[4mm]
&=&-2(\Delta C_{A_n}\otimes\1-\1\otimes\Delta C_{A_n}).
\ea
$$
Hence 
\be\label{58}
I+III=\1\otimes\Delta C_{A_n}-\Delta C_{A_n}\otimes\1.
\ee

By Lemma 2 we get
$$
II=\Delta C_{A_n}\otimes\1-\1\otimes\Delta C_{A_n}.
$$
Therefore
$$
({\cal H})_{12}({\cal H})_{23}({\cal H})_{12}-
({\cal H})_{23}({\cal H})_{12}({\cal H})_{23}=
I+II+III=0.
$$
\hfill $\rule{3mm}{3mm}$

Related to the integrable chain model (\ref{55}), there is a Temperley-Lieb (TL) algebraic 
structure in the sense that the model gives a representation of the TL algebra.
An $(L-1)$-state TL algebra is described by the elements
$e_{i}$, $i=1,2,...,L-1$, satisfying the TL algebraic relations \cite{tl},
\begin{equation}\label{tl}
e_ie_{i\pm 1}e_i=e_i\,,~~~~~
e_ie_j=e_je_i\,,~~~{\sf if~} ~\vert i-j\vert \ge 2\,,
\end{equation}
and
\begin{equation}\label{ei2}
e_i^2=\beta e_i\,,
\end{equation}
where $\beta$ is a complex constant and $i=1,2,\cdots ,L-1$.

We suppose that the representation of an $(L-1)$-state TL algebra on an $L$
chain is of the following form,
\begin{equation}\label{ei}
e_i={\bf 1}_{1}\otimes{\bf 1}_{2}\otimes\cdots\otimes{\bf 1}_{i-1}\otimes
E\otimes{\bf 1}_{i+2}\otimes
\cdots\otimes{\bf 1}_{L}\,,
\end{equation}
where $\1$ is the $(n+1)\times (n+1)$ identity matrix as in section 
3.2 and $E$ is a $(n+1)^{2}\times (n+1)^{2}$ matrix. According to
formulae (\ref{ei2}) and (\ref{tl}) $E$ should satisfy
\begin{equation}\label{ee}
E^{2}=\beta E\,.
\end{equation}
\be\label{ey}
\ba{l}
(E\otimes{\bf 1})({\bf 1}\otimes E)(E\otimes{\bf 1})=E\otimes{\bf
1},\\[3mm]
({\bf 1}\otimes E)(E\otimes{\bf 1})({\bf 1}\otimes E)={\bf 1}\otimes E.
\ea
\ee

From the representations of the TL algebra one can construct
integrable chain models (for the construction of the TL algebraic
representations associated with the quantum $A_1$, $B_n$, $C_n$ and
$D_n$ algebras, see \cite{BK}). It is straightforward to check that
for a given representation of the TL algebra of the form
(\ref{ei}) with $E$ satisfying (\ref{ee}) and (\ref{ey}),
$$
{\check{R}}=E+\frac{-\beta\pm\sqrt{\beta^{2}-4}}{2}\,
{\bf 1}\otimes {\bf 1}
$$
is a solution of the QYBE (\ref{37}).
However in general for a given solution ${\check{R}}$ of the QYBE
(\ref{37}), there does not necessarily exist a TL algebraic 
representation of the form (\ref{ei}) with $E=a{\check{R}}+b$ satisfying
(\ref{ee}) and (\ref{ey}) for any constants $a$ and $b$. Nevertheless
the solutions ${\cal H}$ of the QYBE in our $A_n$ symmetric integrable
model (\ref{55}) do give rise to TL algebraic representations in the
following sense:

{\sf [Theorem 4]}. The following $(n+1)^2\times (n+1)^2$ matrix
\be\label{ae}
E=-\frac{{\cal H}}{n+1}+\1\otimes\1
\ee
gives the $(L-1)$-state TL algebraic representation (\ref{ei}) with
$\beta=2$.

{\sf [Proof]}. What we should check is that $E$ in (\ref{ae})
satisfies equations (\ref{ee}) and (\ref{ey}). By Lemma 1 we have
$$
E^2=(-\displaystyle\frac{{\cal H}}{n+1}+\1\otimes\1)^2
=\displaystyle\frac{(\Delta C_{A_n})^2-2n \Delta C_{A_n}+n^2\1\otimes\1}{(n+1)^2}
=\beta E=2E,
$$
i.e., $\beta=2$.

From Lemma 1 and (\ref{l21}) in Lemma 2 we get
$$
\ba{l}
(E\otimes{\bf 1})({\bf 1}\otimes E)(E\otimes{\bf 1})\\[4mm]
=\displaystyle\frac{-1}{(n+1)^3}[
(\Delta C_{A_n}\otimes\1)(\1\otimes\Delta C_{A_n})(\Delta C_{A_n}\otimes\1)\\[4mm]
~~~-n((\Delta C_{A_n}\otimes\1)(\1\otimes\Delta C_{A_n})+
(\1\otimes\Delta C_{A_n})(\Delta C_{A_n}\otimes\1))\\[4mm]
~~~+2n(n+1)\Delta C_{A_n}\otimes\1
+n^2\1\otimes\Delta C_{A_n}-2(n^3+n^2)\1\otimes\1\otimes\1]\\[4mm]
=\displaystyle\frac{-1}{(n+1)^3}[(n+1)^2\Delta C_{A_n}\otimes\1
-n(n+1)^2\1\otimes\1\otimes\1]=E\otimes{\bf 1}.
\ea
$$

By using Lemma 1 and formula (\ref{l22}) in Lemma 2 we conclude that
$$
\ba{l}
({\bf 1}\otimes E)(E\otimes{\bf 1})({\bf 1}\otimes E)\\[4mm]
=\displaystyle\frac{-1}{(n+1)^3}[
(\Delta C_{A_n}\otimes\1)(\1\otimes\Delta C_{A_n})(\Delta C_{A_n}\otimes\1)\\[4mm]
~~~-n((\Delta C_{A_n}\otimes\1)(\1\otimes\Delta C_{A_n})+
(\1\otimes\Delta C_{A_n})(\Delta C_{A_n}\otimes\1))\\[4mm]
~~~+2n(n+1)\1\otimes\Delta C_{A_n}
+n^2\Delta C_{A_n}\otimes\1-2(n^3+n^2)\1\otimes\1\otimes\1]\\[4mm]
=\displaystyle\frac{-1}{(n+1)^3}[(n+1)^2\1\otimes\Delta C_{A_n}
-n(n+1)^2\1\otimes\1\otimes\1]=\1\otimes E.
\ea
$$
\hfill $\rule{3mm}{3mm}$

From (\ref{ae}) we see that the Hamiltonian of the $A_n$ symmetric
integrable chain model (\ref{55}) can be expressed by the TL algebraic
elements
\be
H_{A_n}=\displaystyle\sum_{i=1}^{L-1}({\cal H})_{i,i+1}
=\displaystyle\sum_{i=1}^{L-1}(n+1)e_i+(n+1)(L-1),
\ee
with $e_i$ as in (\ref{ei}) and $E$ as in (\ref{ae}). Hence instead of the algebraic
Bethe Ansatz method, the energy spectrum of $H_{A_n}$ can also be
studied by using the properties of the TL algebra \cite{levy} (for the
case of Heisenberg spin chain model, $n=1$, see \cite{mahou}). 

\subsection{\bf Integrable $SU(2)$-Symmetric Ladder Models}

We consider ladder models ($M=2$) with $SU(2)$ symmetry. Let $S_i$, $i=1,2,3$, and $C$ 
be the generators of the algebra $SU(2)$ and the Casimir operator respectively.
The coproduct of the algebra is given by $\Delta S_i=\1\otimes S_i +S_i\otimes \1$,
$i=1,2,3$. Accounting to that $\Delta_i^j \Fb (e)=\Fb(\Delta_i^j e)$, $i=1,2,3$,
$j=1,2$, $\forall\, e \in SU(2)$, the generic $h$ is of the form
$\Fb(C_1,C_2,C_3)$, where 
$$
\ba{l}
C_1=\displaystyle\sum_{i=1}^{3}(
S_i \otimes \1 \otimes \1 \otimes S_i+ \1 \otimes S_i \otimes \1 \otimes S_i 
+\1 \otimes \1 \otimes S_i \otimes S_i),\\
C_2=\displaystyle\sum_{i=1}^{3}(
S_i \otimes \1 \otimes \1 \otimes S_i+ S_i \otimes S_i \otimes \1 \otimes \1 
+S_i \otimes \1 \otimes S_i \otimes \1),\\
C_3=\displaystyle\sum_{i=1}^{3}(
S_i \otimes \1 \otimes \1 \otimes S_i+ \1 \otimes S_i \otimes \1 \otimes S_i 
+S_i \otimes \1 \otimes S_i \otimes \1 + \1 \otimes S_i \otimes S_i \otimes \1).
\ea
$$ 

In the spin-$\frac{1}{2}$ representation of the algebra $SU(2)$, the solutions of
the QYBE (\ref{39}) are $16\times 16$ matrices. For instance, it is direct to check that
\be\ba{rcl}
{\cal H}_0&=&\frac{ 108d - 55f}{108} C_{111} 
+\frac{ -72 d + 104 f} {288} C_{112} 
+\frac{ -486 d + 211 f}{270} C_{113} + \frac{ -756 d + 370 f} {216}C_{121}\\[4mm]
&& 
- \frac{29 f}{108} C_{122} 
+ \frac{ 90 d - 31 f}{36} C_{123} + \frac{ 2 d - f}{2} C_{131} 
+ \frac{ -54 d + 26 f}{108}C_{132}\\[4mm]
&& + \frac{ -108 d + 43 f}{540}C_{133}
+ \frac{ -216 d + 80 f}{864}C_{211} 
+ \frac{11 f}{108}C_{212} 
+\frac{ 216 d - 119 f}{108}C_{213}
\ea
\ee
satisfies (\ref{39}) for all $d,f\in\Rb$, 
where $C_{ijk}\equiv C_i\cdot C_j\cdot C_k$, $i,j,k=1,2,3$.

The corresponding solution related to the 
$SU(2)$-symmetric integrable ladder model in \cite{ladder}
can also be expressed in the form $\Fb(C_1,C_2,C_3)$, i.e.,
\be\label{rr0}
\ba{rcl}
{\cal H}&=&-\frac{5}{48}C_{111}
-  \frac{11}{32}C_{112} 
- \frac{61}{30}C_{113} 
 - \frac{41}{48}(C_{121}-C_{122}) + 
\frac{21}{16}C_{123}\\[4mm] 
	  &&+ \frac{3}{4}C_{131} 
-\frac{17}{12}C_{132}+\frac{173}{240}C_{133}
+ \frac{55}{96}C_{211} 
- \frac{5}{3}C_{212} + \frac{131}{48}C_{213}.
\ea
\ee
Through baxterization, ${\cal H}(x)=(x-1){\cal H}+16\, {\rm I}_{16\times 16}$
satisfies the QYBE with spectral parameters:
${\cal H}_{12}(x){\cal H}_{23}(x y){\cal H}_{12}(y)
={\cal H}_{23}(y){\cal H}_{12}(x y){\cal H}_{23}(x)$, where
${\cal H}_{12}(\cdot)={\cal H}(\cdot)\otimes {\rm I}_{4\times 4}$,
${\cal H}_{23}(\cdot)={\rm I}_{4\times 4}\otimes {\cal H}(\cdot)$,
${\rm I}_{n\times n}$ denotes the $n\times n$ identity matrix.
The model can be exactly solved using algebraic Bethe Ansatz method.
It describes a periodic spin ladder system with both isotropic
exchange interactions and biquadratic interactions:
$$
\begin{array}{rcl}
H&=&\displaystyle\frac12
\sum_{i=1}^{L-1}(\frac12+2{\bf S}_{1,i}\cdot{\bf S}_{1,i+1})
(\frac12+2{\bf S}_{2,i}\cdot{\bf S}_{2,i+1})
-\displaystyle\frac12\sum_{i=1}^{L-1}(\frac12+2{\bf S}_{1,i}\cdot{\bf S}_{2,i+1})
(\frac12+2{\bf S}_{2,i}\cdot{\bf S}_{1,i+1})\\[4mm]
&&+\displaystyle\frac{5}{6}\sum_{i=1}^{L-1}
(\frac12+2{\bf S}_{1,i}\cdot{\bf S}_{2,i})
(\frac12+2{\bf S}_{1,i+1}\cdot{\bf S}_{2,i+1}),
\end{array}
$$
where ${\bf S}_{\theta,i}=(\sigma^x_{\theta,i},\sigma^y_{\theta,i},\sigma^z_{\theta,i})/2$, 
$\sigma^x,\sigma^y,\sigma^z$ are Pauli matrices.
${\bf S}_{1,i}$ (resp. ${\bf S}_{2,i}$) is the spin operator on
the first (resp. second) leg of the $i$-th rung of the ladder.

It is also easy to see that for a more general form of (\ref{rr0}),
\be\label{rr}
\ba{rcl}
{\cal H}^\prime&=&\frac{ -45 + 23\,a - 4\,b - 28\,c }{432}
      C_{111} + \frac{ -99 - 3\,a - 3\,b - c }{288}
      C_{112} +  \frac{ -1098 - 91\,a - 118\,b - 16\,c }{540} C_{113}\\[4mm] 
&& + \frac{ -369 - 97\,a - 70\,b + 50\,c }{432} C_{121} + 
\frac{ 396 + 4\,a + 31\,b + 25\,c }{432} C_{122} + 
\frac{ 189 + 29\,a + 20\,b - 4\,c }{144} C_{123}\\[4mm] 
	  &&+ \frac{3}{4}C_{131} + \frac{ -306 - 2\,a - 29\,b - 14\,c }{216}
      C_{132} + \frac{ 1557 - 71\,a + 172\,b + 124\,c }{2160}C_{133}\\[4mm]
	  && + \frac{ 495 - a + 53\,b + 47\,c }{864} C_{211} + 
\frac{ -720 - 22\,a - 49\,b - 43\,c }{432} C_{212} + 
\frac{ 1179 + 91\,a + 118\,b + 16\,c }{432} C_{213}
\ea
\ee
with $a,b,c\in\Rb$, the corresponding ladder model $H^\prime=\displaystyle
\sum_{i=1}^{L-1}{\cal H}^\prime_{i,i+1}$ can also be exactly solved by the
coordinate Bethe ansatz \cite{ladder}.

\section{Integrable Models and Stationary Markov Chains}

\subsection{\bf Stationary Markov Chains}

Let us first briefly recall some concepts of the theory of Markov chains (for a detailed
mathematical description of Markov chains, we refer to \cite{markov}).
Let $\Omega$ denote the sample space (the set of all possible outcomes
of an experiment), which we assume to be
finite or countably infinite. Let $P$ be a probability
measure defined on the $\sigma$-algebra of all subsets of
$\Omega$. Thus any subset $A$ of $\Omega$ is an event with 
probability $P(A)$.

Any function $X\equiv X(\omega)$, $\omega\in \Omega$, 
that maps the sample space into the real numbers 
is then a random variable. A stochastic process is a family
$(X_t)_{t\in I}$, $I$ a certain index set, of random
variables defined on some sample space $\Omega$. If $I$ is countable, i.e.,
$I\in \Nb$, the process is denoted by $X_1,X_2,...$ and
called a discrete-time process. If $I= \Rb_+$, then 
the process is denoted by $\{X_t\}_{t\ge 0}$ and called a
continuous-time process.

The range of $X$ (a subset of real numbers) is called the state space.
In what follows we consider the case where the state space $S$
is countable or finite. In this case the related stochastic process is
called a ({stochastic or random}) {chain}.

Let $E,F$ be two subsets of $\Omega$. We denote by $P(E\vert F)$
the (conditional) probability of $E$ given that
$F$ has occurred. A discrete-time stochastic process $\{X_i\}$,
$i=1,2,...$ with state space $S=\Nb$ is said to satisfy the
Markov property if for every $l$ and all states $i_1,i_2,...,i_l$ it is
(a.s.) true that 
$$
P[X_l=i_l\vert X_{l-1}=i_{l-1},X_{l-2}=i_{l-2},...,X_{1}=i_{1}]=
P[X_l=i_l\vert X_{l-1}=i_{l-1}],
$$
i.e., the values of $X_{l-2},...,X_{1}$ in no way
affect the value of $X_l$, given the value of $X_{l-1}$. 
Such a discrete-time process is
called a {Markov chain}. It is said to be
{stationary} if the probability of going from one state to another is
independent of the time at which the transition is being made. That is, for all
states $i$ and $j$,
$$
P[X_l=j\vert X_{l-1}=i]=P[X_{l+k}=j\vert X_{l+k-1}=i]
$$
for $k=-(l-1),-(l-2),...,-1,0,1,2,...$. In this case we set
$p_{ij}\equiv P[X_l=j\vert X_{l-1}=i]$ and call $p_{ij}$ the 
{transition probability} for going from state $i$ to $j$.

For a discrete time stationary Markov chain $\{X_i\}$, $i\in\Nb$, with a finite
state space $S=\{1,2,3,...,m\}$, there are $m^2$ transition
probabilities $\{p_{ij}\}$, $i,j=1,2,...,m$.  $P=(p_{ij})$ is called the
{transition matrix} corresponding to the discrete-time
stationary Markov
chain $\{X_i\}$. The transition matrix $P$ has the following properties:
\be\label{pm}
p_{ij}\ge 0,~~~\sum_{i=1}^m p_{ij}=1,~~~i,j=1,2,...,m.
\ee
Any square matrix that satisfies condition (\ref{pm}) is called
a {stochastic matrix}.

A continuous-time stochastic process,
$\{X_t\}_{t\in\Rb_+}$ is said to satisfy the Markov property if for all
times $t_0<t_1<...<t_l<t$ and for all $l$ it is true that
$$
P[X_t=j\vert X_{t_0}=i_0,X_{t_1}=i_1,...,X_{t_l}=i_l]
=P[X_t=j\vert X_{t_l}=i_l].
$$
Such a process is called a {continuous-time Markov chain}. It
is said to be stationary if for
every $i$ and $j$ the transition function, $P[X_{t+h}=j\vert X_{t}=i]$,
is independent of $t$. In this case $P(t)=P(X_{t=j}\vert X_0=i)$ is a
semigroup (e.g. on $l^2(S)$), called transition semigroup associated
with the Markov chain. Its generator $Q=(q_{ij})$ has the properties:
\be\label{qm}
q_{ij}\ge 0,~~~i\neq j,~~~~q_{ii}=-\sum_{i\neq j}q_{ij},~~~i,j=1,2,...,m
\ee
and is called an {intensity matrix}. Vice versa, any $Q$ 
(satisfying (\ref{qm}) and properly defined as a closed operator when $S$ is
infinite) gives rise to a unique
continuous-time transition semigroup, $P(t)=e^{Qt}$, $t\geq 0$,
which can be interpreted
as transition semigroup associated to a certain Markov chain
(with state space $S$) \cite{markov}.

The properties of Markov chains are determined by the transition matrix
$P$ resp. intensity matrix $Q$.
If the eigenvalues and eigenstates of $P$ resp. $Q$ are
known, then exact results related to the stochastic processes, such as
time-dependent averages and correlations, can be obtained.

Now we consider a square lattice (in the algebraic sense of sections 2-4)
with $ML$ sites. To every site at the $i$-th rung and $\theta$-th leg 
of the lattice we associate $N$
states described by the variable $\t_{i,\theta}$ taking $N$ integer values,
\be\label{tau}
\t_{i,\theta}\equiv (\t_{i,\theta}^0=0,\t_{i,\theta}^1=1,\t_{i,\theta}^2=2,...,\t_{i,\theta}^N=N).
\ee
We associate to any lattice site a Hilbert space of
dimension $N$ (dim $H_i^\theta =N$). The state space of the algebraic lattice
is then finite and has a total of $(N)^{ML}$ states.

For a given integrable lattice model with Hamiltonian $H$,
the model  remains integrable
if one adds to $H$ a constant term $c$ and multiplies $H$ by a constant
factor $c^\p$. Moreover the eigenvalues of $H$
will not be changed if one changes the local basis, i.e., the following
Hamiltonian, defined by
\be\label{hp}
BHB^{-1},
\ee
where $B=\otimes_{i=1}^{ML}B_i$,
$B_i\equiv {\cal B}$ and ${\cal B}$ is an $N\times N$ non singular matrix, 
has the same eigenvalues as $H$.
Therefore if an integrable lattice model with Hamiltonian $H$ can be
transformed by $B$ (modulo constants $c$, $c^\p$ in $\Cb$)
into a matrix $M=P$ resp. $Q$, in the sense that
\be\label{pmt}
M=B(c^\p H+c\R1)B^{-1},
\ee
where $\R1$ is the $N^{ML}\times (N)^{ML}$ identity matrix,
$B$ as in (\ref{hp}), and $P$ resp. $Q$ as in (\ref{pm}) resp. (\ref{qm}) 
(with $m=(N)^{ML}$),
then $P$ resp. $Q$ defines a discrete-time resp. continuous-time
Markov chain and the related stochastic
process can be studied by using the properties of the corresponding
integrable model with Hamiltonian $H$. 

In the following we discuss the question of whether the integrable lattice
models obtained in the way presented in this paper can
be transformed into some stationary Markov chains through
transformations of the forms (\ref{pmt}). We remark that in general $P$
and $Q$ have different spectra, hence it is necessary to discuss the
cases $M=P$ and $M=Q$ separately.

\smallskip
\subsection{\bf Discrete-time Markov Chains Related to $A_n$
Symmetric Integrable Models}

We first note that for an integrable chain model with Hamiltonian
$H=\sum_{i=1}^{L-1} h_{i,i+1}$ and $(n+1)$ states at every site $i$,
$i=1,2,...,L$, if the sum of the elements in any row of the 
$(n+1)^2\times (n+1)^2$ matrix
$h$ is $1/({L-1})$, then the sum of the elements in any row of 
the matrix $H$ is $1$. Hence if under the following transformation $h\to
h^\p$ given by
\be
h^\p=({\cal B}\otimes {\cal B})(c^\p h+c \1\otimes\1) 
({\cal B}^{-1}\otimes {\cal B}^{-1}),
\ee
the sum of the elements in any row of $h^\p$ is $1/({L-1})$ and
$(h^\p)_{\alpha,\beta}\geq 0$, $\alpha,\beta=1,2,...,(n+1)^2$, for 
some real constants $c^\p, c$ and a non
singular $(n+1)\times (n+1)$ matrix ${\cal B}$, then
$P=\sum_{i=1}^{L-1} h^\p_{i,i+1}$ defines a stationary discrete-time Markov
chain. $P$ has the same eigenvalue spectrum (shifted by a constant)
as the spectrum of the integrable model with Hamiltonian $H$. If $P$ is
invariant under a certain algebra $A$, we call the Markov chain $A$-symmetric.

{\sf [Theorem 5]}. The following matrix
\be\label{pan}
\ba{rcl}
P_{A_n}&=&\displaystyle\frac{1}{(L-1)(n+1)}H_{A_n}
=\displaystyle\frac{1}{(L-1)(n+1)}\sum_{i=1}^{L-1}
(\Delta C_{A_n}+\1\otimes\1)_{i,i+1}\\[4mm]
&=&\displaystyle\frac{1}{(L-1)(n+1)}
\displaystyle\sum_{i=1}^{L-1}\left[(n+1)\displaystyle\sum_{\alpha=1}^{n(n+1)/2}
((e_\alpha)_i(f_\alpha)_{i+1}+(f_\alpha)_i(e_\alpha)_{i+1})\right.\\[5mm]
&&+\displaystyle\sum_{\alpha=1}^{n}\alpha(n+1-\alpha)
(h_\alpha)_i(h_\alpha)_{i+1}\\[5mm]
&&\left.+\displaystyle\sum_{\alpha=1}^{n}\displaystyle\sum_{\beta=1}^{n-\alpha}
\alpha(n+1-\alpha-\beta)((h_\alpha)_i(h_{\alpha+\beta})_{i+1}
+(h_{\alpha+\beta})_i(h_{\alpha})_{i+1})\right]+\displaystyle\frac{1}{(n+1)}
\ea
\ee
defines a stationary discrete-time $A_n$ symmetric Markov chain.

{\sf [Proof]}. I. Set $h^\p\equiv\frac{1}{(L-1)(n+1)}(\Delta
C_{A_n}+\1\otimes\1)$. Then 
\be\label{pform}
P_{A_n}=\sum_{i=1}^{L-1} h^\p_{i,i+1}.
\ee 
From formula (\ref{48}) we have
\be\label{lhp}
\ba{rcl}
(h^\p)_{\alpha\beta}&=&\displaystyle\frac{1}{(L-1)(n+1)}(\Delta
C_{A_n}+\1\otimes\1)_{\alpha\beta}\\[4mm]
&=&\displaystyle\frac{1}{(L-1)(n+1)}
[\delta_{\alpha\beta}[(n+1)\delta_{\alpha,l(n+1)+l+1}-1]\\[4mm]
&&+(n+1)[\delta_{\alpha,j(n+2)+k+2}\delta_{\beta,(j+1)(n+2)+k(n+1)}\\[4mm]
&&+\delta_{\beta,j(n+2)+k+2}\delta_{\alpha,(j+1)(n+2)+k(n+1)}]+\delta_{\alpha\beta}]\\[4mm]
&=&\displaystyle\frac{1}{L-1}
[\delta_{\alpha\beta}\delta_{\alpha,l(n+1)+l+1}
+\delta_{\alpha,j(n+2)+k+2}\delta_{\beta,(j+1)(n+2)+k(n+1)}\\[4mm]
&&+\delta_{\beta,j(n+2)+k+2}\delta_{\alpha,(j+1)(n+2)+k(n+1)}]\geq 0.
\ea
\ee
Therefore $(P_{A_n})_{\alpha\beta}\geq 0$,
$\alpha,\beta=1,2,...,(n+1)^2$.

II. By using the identity (\ref{identity}), we get
$$
\ba{rcl}
\displaystyle\sum_{\beta=1}^{(n+1)^2}(h^\p)_{\alpha\beta}
&=&\displaystyle\frac{1}{(L-1)(n+1)}\displaystyle\sum_{\beta=1}^{(n+1)^2}
[(n+1)\delta_{\alpha\beta}\delta_{\alpha,l(n+1)+l+1}\\[5mm]
&&+(n+1)[\delta_{\alpha,j(n+2)+k+2}\delta_{\beta,(j+1)(n+2)+k(n+1)}\\[4mm]
&&+\delta_{\beta,j(n+2)+k+2}\delta_{\alpha,(j+1)(n+2)+k(n+1)}]]\\[4mm]
&=&\displaystyle\frac{1}{(L-1)(n+1)}(n+1)=\displaystyle\frac{1}{L-1}.
\ea
$$
Hence the sum of the elements of any row of the matrix $P_{A_n}$ is one,
i.e., $\sum_{\beta=1}^{(n+1)^{L}}(P_{A_n})_{\alpha\beta}
=\sum_{\beta=1}^{(n+1)^{L}}(\sum_{i=1}^{L-1} h^\p_{i,i+1})_{\alpha\beta}=1$.

III. As $H_{A_n}$ is invariant under $A_n$,
$P_{A_n}=\frac{H_{A_n}}{(L-1)(n+1)}$ is obviously invariant under $A_n$
and has the same spectrum as $H_{A_n}$.

By the definition (\ref{pm}) $P_{A_n}$ is the transition matrix of a
stationary discrete-time $A_n$ symmetric Markov chain. 
\hfill $\rule{3mm}{3mm}$

The state space of this stationary discrete-time 
$A_n$ symmetric Markov chain associated with the 
stochastic matrix $P_{A_n}$ is
$S=(1,2,...,(n+1)^{L})$, which corresponds to $(n+1)^{L}$
states, 
\be\label{taut}
(\t_0\otimes \t_1\otimes...\otimes \t_n),
\ee
$\t_i\equiv \t_{i,1}$ as in (\ref{tau}) with $\theta=1$, of 
the algebraic chain with $L$ lattice sites. 
This stationary discrete-time $A_n$ symmetric Markov chain model
describes a chain with $L$ sites and $n+1$ possible
states, say, $\t_i^{1},...,\t_i^{n+1}$, at site $i$, $i=1,...,L$. By
calculation the allowed dynamics for any nearest neighbour pair is the
interchange of their states: ($\t_i^{\alpha},\t_{i+1}^{\beta}$)
$\to$ ($\t_i^{\beta},\t_{i+1}^{\alpha}$), $\alpha,\beta\in
\{1,...,n+1\}$. For $n=1$, there are two possible states at every site:
empty or occupied by one particle. The $A_1$ Markov chain describes then
stochastic hopping of particle into left and right vacancies, which is
the well known $SU(2)$ random chain. For $n=2$, i.e. the $SU(3)$ case, there
are three possible states at every site, say, empty, one spin up
and one spin down particle states. This model describes
stochastic hopping of spin up and down particles into left and right
vacancies, interchanging of one spin up and one spin down particle at
the nearest neighbours.

The properties of a Markov chain are determined
by the transition matrix $P=(p_{ij})$. A subset $C$ of 
the state space $S$ is called closed if $p_{ij}=0$ for all $i\in C$ and
$j\not\in C$. If a closed set consists of a single state, then that
state is called an {absorbing state}. A Markov chain is called 
{irreducible}
if there exists no nonempty closed set other than $S$ itself. A non
irreducible Markov chain is said to be {reducible}.

From formula (\ref{lhp}) we have 
\be\label{hpp}
\ba{l}
(h^\p)_{\alpha\alpha}=(h^\p)_{(n+1)^2,(n+1)^2}=\frac{1}{L-1},~~~
(h^\p)_{\alpha\beta}=(h^\p)_{\beta\alpha}=0,~~~\beta\neq\alpha,\\[3mm]
\alpha=l(n+1)+l+1,~~~l=0,1,...,n.
\ea
\ee

Let 
\be\label{s0}
S_0=\left(\alpha\vert\alpha=l\frac{(n+1)((n+1)^{L-1}-
1)+n}{n}+1\right),~~~l=0,1,...,n,
\ee
be a subset of the state space $S$.
From formula (\ref{pform}), with 
$$
(h^\p)_{i,i+1}=\1_1\otimes\1_2\otimes...\otimes\1_{i-1}\otimes
h^\p\otimes\1_{i+2}\otimes...\otimes\1_{L+1},
$$
we get
\be\label{panp}
(P_{A_n})_{\alpha\alpha}=1,~~~
(P_{A_n})_{\beta\alpha}=(P_{A_n})_{\alpha\beta}=0,~~~\beta\neq\alpha,
~~~\alpha\in S_0.
\ee

Therefore the $n+1$ states in $S_0$ are absorbing states of the
Markov chain $P_{A_n}$. This chain is by definition reducible. For a
reducible Markov chain the ``long time" probability distribution, if it
exists, may depend on the initial conditions, i.e.,
$\lim_{l\to\infty}(P_{A_n})^{l}_{\gamma\beta}$ may depend on $\gamma$.
From the properties (\ref{panp}) of $P_{A_n}$, we see that if the Markov
chain $P_{A_n}$ is initially in one of the states $\alpha\in S_0$, it
will remain in that state $\alpha$ forever. These $n+1$ absorbing states 
correspond to the states of the algebraic chain through (\ref{taut}).
For instance, the states $1$ and $(n+1)^{L}$ in $S$ correspond to the
states $(0,0,...,0)$ (all the
sites of the algebraic chain are at state $0$) and $(n,n,...,n)$ (all the
sites of the algebraic chain are at state $n$).

\subsection{\bf Continuous-time Markov Chains Related to $A_n$
Symmetric Integrable Models}

For an integrable chain model with Hamiltonian 
$H=\sum_{i=1}^{L-1} h_{i,i+1}$ and with $(n+1)$ states at every site of the chain, 
if the sum of the elements in any column of the matrix
$h$ is $0$, the sum of the elements in any column of 
the matrix $H$ is also $0$. Hence if under the following transformation
$h\to h^{\p\p}$ with:
\be
h^\pp=({\cal B}\otimes {\cal B})(c^\p h+c \1\otimes\1)({\cal B}^{-1}\otimes {\cal B}^{-1}),
\ee
the sum of the elements in any column of $h^\pp$ is $0$ and
$(h^\pp)_{\alpha,\beta}\geq 0$, $\alpha\neq\beta=1,2,...,(n+1)^2$, for
some real constants $c^\p, c$ and a non
singular $(n+1)\times (n+1)$ matrix ${\cal B}$, then
$Q=\sum_{i=1}^{L-1} h^\pp_{i,i+1}$ is the intensity matrix for some stationary 
continuous-time Markov
chain. $Q$ has the same eigenvalue spectrum (shifted by a constant)
as the spectrum of the Hamiltonian $H$. We call the Markov chain 
{$A$ symmetric} if $Q$ is invariant under the algebra $A$.

{\sf [Theorem 6]}. The following matrix $Q$ is the intensity matrix of a
stationary continuous-time Markov chain:
\be\label{qan}
\ba{rcl}
Q_{A_n}&=&H_{A_n}-(n+1)(L-1)
=\displaystyle\sum_{i=1}^{L-1}(\Delta C_{A_n}-n\1\otimes\1)_{i,i+1}\\[4mm]
&=&\displaystyle\sum_{i=1}^{L-1}\left[(n+1)\displaystyle\sum_{\alpha=1}^{n(n+1)/2}
((e_\alpha)_i(f_\alpha)_{i+1}+(f_\alpha)_i(e_\alpha)_{i+1})
+\displaystyle\sum_{\alpha=1}^{n}\alpha(n+1-\alpha)
(h_\alpha)_i(h_\alpha)_{i+1}\right.\\[5mm]
&&\left.+\displaystyle\sum_{\alpha=1}^{n}\displaystyle\sum_{\beta=1}^{n-\alpha}
\alpha(n+1-\alpha-\beta)((h_\alpha)_i(h_{\alpha+\beta})_{i+1}
+(h_{\alpha+\beta})_i(h_{\alpha})_{i+1})\right]-n(L-1).
\ea
\ee

{\sf [Proof]}. Set $h^\pp=(\Delta C_{A_n}-n\1\otimes\1)$. Then 
\be\label{qanhpp}
Q_{A_n}=\sum_{i=1}^{L-1} h^\pp_{i,i+1}.
\ee
From (\ref{48}) we observe that, for $\alpha\neq\beta$,
$$
\ba{ll}
h^\pp_{\alpha\neq\beta}=
&(n+1)[\delta_{\alpha,j(n+2)+k+2}\delta_{\beta,(j+1)(n+2)+k(n+1)}\\[3mm]
&+\delta_{\beta,j(n+2)+k+2}\delta_{\alpha,(j+1)(n+2)+k(n+1)}]\geq 0.
\ea
$$
Therefore $(Q_{A_n})_{\alpha\neq\beta}\geq 0$, $\alpha,\beta=1,2,...,
(n+1)^{L}$.

Again by (\ref{48}) the sum of the elements in any given column $\beta$
of the matrix $h^\pp$ is
$$
\ba{l}
\displaystyle\sum_{\alpha=1}^{(n+1)^2}
h^\pp_{\alpha\beta}=\displaystyle\sum_{\alpha=1}^{(n+1)^2}(n+1)
[\delta_{\alpha\beta}(\delta_{\alpha,l(n+1)+l+1}-1)\\[5mm]
~~~+(n+1)[\delta_{\alpha,j(n+2)+k+2}\delta_{\beta,(j+1)(n+2)+k(n+1)}
+\delta_{\beta,j(n+2)+k+2}\delta_{\alpha,(j+1)(n+2)+k(n+1)}]\\[3mm]
=\displaystyle\sum_{\alpha\neq l(n+1)+l+1}^{(n+1)^2}(n+1)
[-\delta_{\alpha\beta}
+\delta_{\alpha,j(n+2)+k+2}\delta_{\beta,(j+1)(n+2)+k(n+1)}\\[5mm]
~~~+\delta_{\beta,j(n+2)+k+2}\delta_{\alpha,(j+1)(n+2)+k(n+1)}]\\[3mm]
=(n+1)(-1+\delta_{\beta,(j+1)(n+2)+k(n+1)}\vert_{\alpha=j(n+2)+k+2}
+\delta_{\beta,j(n+2)+k+2}\vert_{\alpha=(j+1)(n+2)+k(n+1)})\\[3mm]
=0,
\ea
$$
i.e., the sum of the elements in any given column $\beta$,
$\beta=1,2,...,(n+1)^2$, of the matrix $h^\pp$ is zero.
Therefore the sum of the elements in any given column $\beta$,
$\beta=1,2,...,(n+1)^{L}$, of the matrix $Q_{A_n}$ is also zero,
$\sum_{\alpha=1}^{(n+1)^{L}}(Q_{A_n})_{\alpha\beta}=0$.
At last $Q_{A_n}=H_{A_n}-(n+1)(L-1)$ is obviously $A_n$ symmetric with the
same spectrum (shifted by a constant) as $H_{A_n}$. \hfill $\rule{3mm}{3mm}$

The long run distribution of the Markov chain described in Theorem 6
is given by the vector ${\bf \pi}=(\pi_1,\pi_2,...)$, where
$\pi_i$ represents the ``long time" probability of the state $i\in S$, 
satisfying
\be\label{pi}
\sum_{\alpha=1}^{(n+1)^{L}} (Q_{A_n})_{\alpha\beta}\pi_\alpha=0,
~~~\forall\beta\in S,~~~\sum_{\alpha=1}^{(n+1)^{L}}\pi_\alpha=1.
\ee
However as this Markov chain is reducible, the solution of the
equation (\ref{pi}) is not unique but depends on the initial conditions.
From (\ref{48}) we see that
$$
(h^\pp)_{\alpha\beta}=(h^\pp)_{\beta\alpha}=0,~~~\forall \beta,~~~~
\alpha=l(n+1)+l+1,~~~l=0,1,...,n.
$$
Hence from (\ref{qanhpp}) we get
$$
(Q_{A_n})_{\alpha\beta}=(Q_{A_n})_{\beta\alpha}=0,~~~\alpha\in S_0,~~
\forall \beta,
$$
with $S_0$ as in (\ref{s0}).
Therefore if this Markov chain is initially in a given state $\alpha\in
S_0$, it will remain in that state. 

The states $\beta\not\in S_0$ form a closed subset of $S$.
From (\ref{48}) and (\ref{qanhpp}) one also learns that the absolute value
of all the nonzero elements of any column of the intensity 
matrix $Q_{A_n}$ are equal. Let $S^\p$ be a closed subset of $S$ 
with $l$ elements. If the Markov chain is initially in the closed set
$S^\p$, then it will remain in $S^\p$ and the long run distribution
is ${\bf \pi}=(\pi_1,\pi_2,...,\pi_{(n+1)^{L+1}})$, where $\pi_i=1/l$
for $i\in S^\p$ and $\pi_i=0$ if $i\not\in S^\p$.

\smallskip
\subsection{\bf Discrete and Continuous-time Markov Chains Related to $SU(2)$
Symmetric Integrable Ladder Models}
 
To every site on the $i$-th rung and $\theta$-th leg, $\theta=1,2$,
of the ladder we associate
states described by the variable $\t_{i,\theta}$ taking values $0$ and $1$.
The state space of this algebraic ladder 
is then finite and has a total of $m=2^{2L}$ states.

{\sf [Theorem 7]}. The following matrix
\be\label{panladd}
P_{SU(2)}=\displaystyle\frac{1}{4(L-1)(18 + 4a + 4b + c)}
\sum_{i=1}^{L-1}{\cal H}^{\prime\prime}_{i,i+1},
\ee
defines a stationary discrete-time $SU(2)$-symmetric integrable Markov ladder for
$a+2b\geq 16$.
The operator ${\cal H}^{\prime\prime}$ is given by
\be\label{hppladd}
{\cal H}^{\prime\prime}=
\left(
\ba{cccccccccccccccc}
  a_1& a_2& a_2& a_2& a_3& a_4& a_4& a_4& a_3& a_4& a_4& a_4& a_3& a_4& a_4& a_4\\[3mm] 
  a_2& a_5& a_6& a_6& a_7& a_3& a_8& a_8& a_8& a_9& a_4& a_4& a_8& a_9& a_4& a_4\\[3mm] 
  a_2& a_6& a_5& a_6& a_8& a_4& a_9& a_4& a_7& a_8& a_3& a_8& a_8& a_4& a_9& a_4\\[3mm] 
  a_2& a_6& a_6& a_5& a_8& a_4& a_4& a_9& a_8& a_4& a_4& a_9& a_7& a_8& a_8& a_3\\[3mm] 
  a_3& a_7& a_8& a_8& a_5& a_2& a_6& a_6& a_9& a_8& a_4& a_4& a_9& a_8& a_4& a_4\\[3mm] 
  a_4& a_3& a_4& a_4& a_2& a_1& a_2& a_2& a_4& a_3& a_4& a_4& a_4& a_3& a_4& a_4\\[3mm] 
  a_4& a_8& a_9& a_4& a_6& a_2& a_5& a_6& a_8& a_7& a_3& a_8& a_4& a_8& a_9& a_4\\[3mm] 
  a_4& a_8& a_4& a_9& a_6& a_2& a_6& a_5& a_4& a_8& a_4& a_9& a_8& a_7& a_8& a_3\\[3mm] 
  a_3& a_8& a_7& a_8& a_9& a_4& a_8& a_4& a_5& a_6& a_2& a_6& a_9& a_4& a_8& a_4\\[3mm] 
  a_4& a_9& a_8& a_4& a_8& a_3& a_7& a_8& a_6& a_5& a_2& a_6& a_4& a_9& a_8& a_4\\[3mm] 
  a_4& a_4& a_3& a_4& a_4& a_4& a_3& a_4& a_2& a_2& a_1& a_2& a_4& a_4& a_3& a_4\\[3mm] 
  a_4& a_4& a_8& a_9& a_4& a_4& a_8& a_9& a_6& a_6& a_2& a_5& a_8& a_8& a_7& a_3\\[3mm] 
  a_3& a_8& a_8& a_7& a_9& a_4& a_4& a_8& a_9& a_4& a_4& a_8& a_5& a_6& a_6& a_2\\[3mm] 
  a_4& a_9& a_4& a_8& a_8& a_3& a_8& a_7& a_4& a_9& a_4& a_8& a_6& a_5& a_6& a_2\\[3mm] 
  a_4& a_4& a_9& a_8& a_4& a_4& a_9& a_8& a_8& a_8& a_3& a_7& a_6& a_6& a_5& a_2\\[3mm] 
  a_4& a_4& a_4& a_3& a_4& a_4& a_4& a_3& a_4& a_4& a_4& a_3& a_2& a_2& a_2& a_1
\ea
\right)
\ee
where $a_1=66 + a + 4b + 4c$, $a_2=-10 + a + 2b$, $a_3=6 + a + 2b$,
$a_4=2 + a$, $a_5=54 + a + 4b + 4c$, $a_6=-16 + a + 2b$,
$a_7=14 + a$, $a_8=8 + a$, $a_9=a + 2b$. ${\cal H}^{\prime\prime}_{i,i+1}$
acts on the $i$ and $i+1$ rungs as defined in (\ref{pp}).

{[\sf Proof].} For the integrable ladder model (\ref{rr}) with Hamiltonian 
$H^\prime=\displaystyle\sum_{i=1}^{L-1}{\cal H}^{\prime}_{i,i+1}$, 
the system  remains integrable under (\ref{hp}).
It is straightforward to prove that ${\cal H}^{\prime\prime}
={\cal B}{\cal H}^{\prime}{\cal B}^{-1}$, where
$$
{\cal B}=\left(
\ba{cccc}
-1 & 1 & 0 & 0\\[3mm]
1 & 1/2 & -1/2 & 1\\[3mm]
0 & -1/2 & -3/2 & 0\\[3mm]
0 & 1 & 0 & -1
\ea
\right).
$$
Therefore the Hamiltonian systems $H^{\p}$ and 
$H^{\p\p}=\displaystyle
\sum_{i=1}^{L-1}{\cal H}^{\prime\prime}_{i,i+1}$ satisfy the relation (\ref{hp})
with $B_i={\cal B}$, $i=1,2,...,L$. Hence $H^{\p\p}$ is also $SU(2)$-symmetric and integrable
with the same spectrum as $H^{\p}$.

For $a+2b \geq 0$, as the entries of ${\cal H}^{\prime\prime}$ are positive,
$H^{\prime\prime}_{\alpha\beta}\geq 0$, $\alpha,\beta=1,2,...,2^{2L}$. From (\ref{hppladd}) we
also have $\sum_{\alpha=1}^{16}{\cal H}^{\p\p}_{\alpha\beta}=4(18+4a+4b+c)$, $\forall\, \beta=
1,2,...,16$. By the definition (\ref{pm}) $P_{SU(2)}$ is the transition matrix
of a stationary discrete-time $SU(2)$-symmetric integrable Markov processes.
\hfill $\rule{3mm}{3mm}$

The state space of this Markov processes associated with the 
stochastic matrix $P_{SU(2)}$ is $S=(1,2,...,2^{2L})$.
Generally there is no closed subset $C$ of
the state space $S$ such that $(P_{SU(2)})_{ij}=0$ for all $i\in C$ and
$j\not\in C$. However in certain parameter regions for $a,b,c$,
from (\ref{hppladd}) one can see that there can exist such closed subsets
$C$ of $S$ (the Markov processes are by definition reducible in these cases). 
From (\ref{hppladd})
it can also be seen that there exists no absorbing state for this Markov process.

By using the results used in the proof of theorem 7, we have also
the following integrable stationary continuous-time Markov process:

{\sf [Theorem 8]}. The matrix 
\be\label{qanladd}
Q_{SU(2)}=H^{\p\p}-4(L-1)(18 + 4a + 4b + c)=
\sum_{i=1}^{L-1}({\cal H}^{\p\p}-4(18 + 4a + 4b + c))_{i,i+1}
\ee
is the intensity matrix of a stationary continuous-time Markov process.

\section{Conclusion and Remark}

Using the Casimir operators and coproduct operations of algebras, we have
given a general way to construct square lattice models with a certain 
Lie or quantum Lie algebraic symmetry. As applications we discussed integrable 
$A_n$ symmetric chain models and $SU(2)$ invariant ladder models.
We have shown that the stochastic processes
correspond to both $A_n$ symmetric integrable chain
models and $SU(2)$ invariant ladder models
are exactly solvable stationary discrete-time (resp.
continuous-time) Markov chains with the
transition matrices (resp. intensity matrices) which coincide with those
of the corresponding integrable models.
Other symmetric integrable lattice models ( e.g. with $B_n$,
$C_n$, $D_n$ symmetry) and integrable Markov models can be investigated in a
similar way.

\end{document}